# Characterization of the second- and third-order nonlinear optical susceptibilities of monolayer $MoS_2$ using multiphoton microscopy

**R I Woodward**[1], **R T Murray**[1], **C F Phelan**[2], **R E P de Oliveira**[2], **T H Runcorn**[1], **E J R Kelleher**[1], **S Li**[3], **E C de Oliveira**[2], **G J M Fechine**[2], **G Eda**[3] and **C J S de Matos**[2]

[1] Femtosecond Optics Group, Department of Physics, Imperial College London, London, UK
[2] MackGraphe–Graphene and Nanomaterials Research Center, Mackenzie Presbyterian University, São Paulo, Brazil
[3] Centre for Advanced 2D Materials, National University of Singapore, Singapore

E-mail: cjsdematos@mackenzie.br

**Abstract.** We report second- and third-harmonic generation in monolayer $MoS_2$ as a tool for imaging and accurately characterizing the material's nonlinear optical properties under 1560 nm excitation. Using a surface nonlinear optics treatment, we derive expressions relating experimental measurements to second- and third-order nonlinear sheet susceptibility magnitudes, obtaining values of $|\chi_s^{(2)}| = 2.0 \times 10^{-20}$ m$^2$ V$^{-1}$ and for the first time for monolayer $MoS_2$, $|\chi_s^{(3)}| = 1.7 \times 10^{-28}$ m$^3$ V$^{-2}$. These sheet susceptibilities correspond to effective bulk nonlinear susceptibility values of $|\chi_b^{(2)}| = 2.9 \times 10^{-11}$ m V$^{-1}$ and $|\chi_b^{(3)}| = 2.4 \times 10^{-19}$ m$^2$ V$^{-2}$, accounting for the sheet thickness. Experimental comparisons between $MoS_2$ and graphene are also performed, demonstrating ∼3.4 times stronger third-order sheet nonlinearity in monolayer $MoS_2$, highlighting the material's potential for nonlinear photonics in the telecommunications C band.

## 1. Introduction

Two-dimensional (2D) materials are attracting significant interest due to their unprecedented optical and electronic properties. While graphene remains the most widely studied 2D material, many other monolayer and few-layer atomic crystals possessing distinct yet complementary properties have recently been discovered [1, 2]. In particular, semiconducting few-layer transition metal dichalcogenides (TMDs), such as molybdenum disulfide (MoS$_2$), have received much attention. Few-layer MoS$_2$ exhibits ultrafast carrier dynamics, strong photoluminescence, saturable absorption and a bandgap which can be tuned by varying the number of atomic layers (from a 1.3 eV indirect gap for bulk MoS$_2$ to a direct 1.9 eV gap for a monolayer) [3–6]. These outstanding characteristics suggest the material has great potential as a platform for developing next-generation electronic, optoelectronic and photonic technologies, including transistors with current on/off ratios exceeding $10^8$, ultrashort pulse lasers, flexible sensors and valleytronic devices [7–10].

As the catalogue of 2D materials continues to grow, an increasing need exists for a thorough and comparative characterization of their properties and performance. Nonlinear microscopy—a general term used to describe any microscopy technique that exploits a nonlinear optical interaction, including harmonic generation, four-wave mixing, and multiphoton absorption—has been demonstrated as a powerful tool for imaging and characterization of 2D atomic crystals [11–22]. Second harmonic generation (SHG) has been observed in monolayer and few-layer MoS$_2$ [16–20], and has been used to probe the crystal symmetry [18] and grain orientations [19] of fabricated samples. This technique, however, is limited to samples with an odd number of layers, as both bulk and even-layer-count few-layer crystals exhibit inversion symmetry; thus, second-order nonlinear effects are electric dipole forbidden. An attractive alternative is to harness third-harmonic generation (THG), which occurs irrespective of inversion symmetry [12, 23, 24]. Wang *et al.* recently reported THG from MoS$_2$ thin films of 7–15 atomic layers [21], suggesting THG could provide complementary information in multiphoton microscopy. Such a high layer count is approaching the bulk regime [1], however, and the technique has yet to be extended to single-layer MoS$_2$.

In addition to being a tool for crystal characterization, SHG and THG imaging are important techniques for evaluating fundamental material parameters, such as the nonlinear optical susceptibility tensors $\chi^{(2)}$ and $\chi^{(3)}$ that determine the strength of nonlinear processes, including the Pockels and Kerr effects, polarization rotation, frequency conversion, and phase conjugation—all of which define the usefulness of a material as a platform for the development of optical devices. Thus, it is crucial to characterize the nonlinearity of 2D materials, in particular at technologically relevant wavelengths, such as the telecommunications C band (1530-1565 nm), where emerging semiconductor materials could have major impact for on-chip switching and signal processing.

To relate experimental measurements to the magnitude of nonlinear susceptibility tensors, the 2D nature of monolayer atomic crystals must be considered. A variety of different formalisms have been adopted in literature to date to account for infinitesimally thin materials, leading to a wide variation in reported material properties: published values for $|\chi^{(3)}|$ in graphene, for example, vary by six orders of magnitude [25]. Further work is therefore needed to determine appropriate figures of merit for describing the nonlinear optical response of emerging 2D materials and to compare their performance.

Here, we determine the magnitude of the second- and third-order nonlinearity susceptibilities in monolayer MoS$_2$ using a power-calibrated multiphoton microscope setup by treating the 2D material as a nonlinear polarization sheet, adopting and extending established work on surface nonlinear optics [26]. We also characterize monolayer graphene, enabling a direct experimental comparison that shows MoS$_2$ possesses a stronger third-order nonlinear response and hence, could be more promising for practical nonlinear photonic applications.

## 2. Methods

First, monolayer MoS$_2$ flakes are fabricated by chemical vapor deposition (CVD) on a silicon (Si) substrate with a ∼300 nm silica (SiO$_2$) coating layer, as described in Ref. [27]. Atomic force microscopy (AFM) and Raman microscopy are used to identify and characterize single-layer flakes [Figure 1(a)-(b)], showing the expected ∼0.7 nm thickness for a monolayer on the substrate and separation of ∼19.4 cm$^{-1}$ between the $E^1_{2g}$ and $A_{1g}$ Raman modes [28].

A microscope setup is developed to enable linear optical imaging using a green LED source and CCD camera in addition to measurement of harmonics that are generated when the sample is excited at normal incidence by a 1560 nm mode-locked Er:fiber laser (Figure 2). Pump pulses with 150 fs duration at 89 MHz repetition rate are focussed through a 20× objective lens (0.50 NA) to a $1/e^2$ diameter of 3.6 μm (with Rayleigh range ∼ 6.5 μm). Pump light is linearly-polarized and a half-waveplate (HWP) is included to control the incident polarization. Reflected harmonics can be observed overlaid on the linear optical image to identify the position of the pump light on the sample [Figure 1(c)] or measured on a spectrometer. The sample is mounted on a piezo-controlled triaxial translation stage, enabling automated raster scanning across the material to construct the nonlinear images.

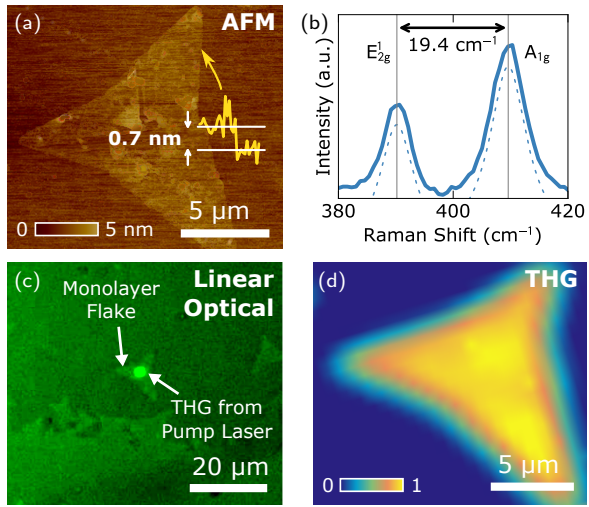

**Figure 1.** Characterization of monolayer MoS$_2$ flake on Si/SiO$_2$ substrate: (a) AFM image and height profile inset; (b) Raman spectrum [vertical lines show the peak positions, obtained by Lorentzian fitting (dashed lines)]; (c) optical image (with the monolayer and focussed pump beam position highlighted); (d) THG image.

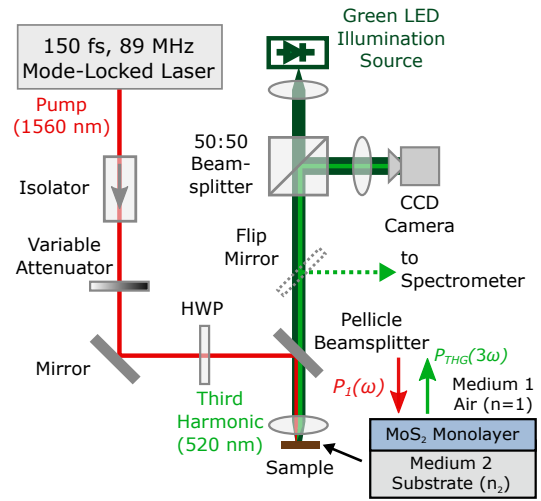

**Figure 2.** Experimental microscope setup for simultaneous linear and nonlinear optical imaging [the second harmonic was also generated (not shown), following the same path as the third-harmonic].

To relate measured intensity values using the spectrometer to the power at the sample, the system is carefully calibrated. The wavelength- and polarization-dependent transmission factors of all components are characterized using a white-light source, laser diode and polarizers, and accounted for in subsequent measurements. Finally, to verify the setup for quantifying nonlinear frequency conversion, the response of ZnS, a well-known bulk material, is measured, from which we obtain second- and third-order susceptibility values in good agreement with literature (see Supplementary Information).

## 3. Results

### 3.1. MoS$_2$ Characterization

Second-harmonic (at 780 nm) and third-harmonic (at 520 nm) signals are clearly observed from monolayer MoS$_2$ flakes for an incident peak intensity of $\sim 10^{14}$ W m$^{-2}$ [Figure 3(a)-(b)]. The sample geometry is imaged by raster scanning the pump beam position and recording the THG intensity [Figure 1(d)], producing a higher contrast image than is possible with the linear optical microscopy part of the setup [Figure 1(c)]. We note that a similar image of monolayer MoS$_2$ could be obtained by recording the SHG intensity [16, 17], although the benefit of THG microscopy is that the technique is widely applicable to 2D materials with any number of layers, in addition to providing higher spatial resolution.

To quantify the nonlinear response of monolayer MoS$_2$, the modulus of the nonlinear susceptibility can be extracted from measurements of the intensity of generated harmonics compared to the pump. For this calculation we follow the theoretical surface SHG formalism of Shen [26]. Here, a surface is treated as a sheet of dipoles radiating coherently and nonlinearly, with a distinct dielectric constant and nonlinear susceptibility to the two materials meeting at the interface. Thus, the second-order nonlinear response of a 2D material is quantified by a nonlinear sheet susceptibility $|\chi_s^{(2)}|$ [17]. Local-field correction factors (i.e. Fresnel reflection coefficients) are also included to account for the boundary conditions. This approach is well suited to analysis of nonlinear optics in 2D materials where the infinitesimally small thickness not only indicates that no phase matching conditions apply along the direction normal to the sheet (and thus, to normally incident light), but also leads to nonlinearly radiated waves in both forwards and backwards directions. This latter feature cannot be obtained from a simple bulk nonlinear optics treatment.

In this work we apply this theory to monolayer MoS$_2$, treated as a nonlinear sheet at the interface between air and the dielectric substrate (Figure 2), and expand the sheet polarization susceptibility formalism to THG in order to compute $|\chi_s^{(3)}|$. Our derivation (see Supplementary Material) considers light at normal incidence to the sample and assumes negligible contribution from the nonlinearity of air or substrate, that the index of air is 1 and that the substrate dispersion is negligible (we also calculated susceptibility values including the effect of dispersion, obtaining $< 0.8\%$ difference, verifying this assumption is a valid simplification). SI units are used throughout. We find:

$$I_{\text{SHG}}(2\omega) = \frac{1}{\epsilon_0}\left[\frac{1}{2c}\left(\frac{2}{1+n_2}\right)^2\right]^3 (2\omega)^2 |\chi_s^{(2)}|^2 I_1^2(\omega) \quad (1)$$

and

$$I_{\text{THG}}(3\omega) = \frac{1}{\epsilon_0^2}\left[\frac{1}{2c}\left(\frac{2}{1+n_2}\right)^2\right]^4 (3\omega)^2 |\chi_s^{(3)}|^2 I_1^3(\omega) \quad (2)$$

where $c$ is the speed of light in vacuum, $\epsilon_0$ is the permittivity of free-space, $n_2 \sim 1.5$ is the substrate index, $\omega$ is the pump angular frequency, $I_1(\omega)$ is the focussed pump peak intensity in air, $|\chi_s^{(2)}|$ and $|\chi_s^{(3)}|$ are the magnitudes of the sheet susceptibility for second- and third-order nonlinearity, respectively. We relate peak intensities to experimentally measured time-averaged power values assuming Gaussian-shaped pulses and Gaussian beam optics, including correction factors to account for the pulse shortening and spot size reduction of the harmonics compared to the pump (see Supplementary Material):

$$P_{\text{SHG}}(2\omega) = \frac{16\sqrt{2}S|\chi_s^{(2)}|^2\omega^2}{c^3\epsilon_0 f\pi r^2 t_{\text{fwhm}}(1+n_2)^6} P_1^2(\omega) \quad (3)$$

and

$$P_{\text{THG}}(3\omega) = \frac{64\sqrt{3}S^2|\chi_s^{(3)}|^2\omega^2}{c^4\epsilon_0^2 (ft_{\text{fwhm}}\pi r^2)^2(1+n_2)^8} P_1^3(\omega) \quad (4)$$

where $f$ is the pump laser repetition rate, $S = 0.94$ is a shape factor for Gaussian pulses, $t_{\text{fwhm}}$ is the pulse full width at half maximum, and $P_1(\omega)$ is the average pump power.

An Si substrate with $\sim 300$ nm $SiO_2$ overlayer is commonly chosen for 2D transition metal dichalcogenide crystal growth and inspection as it facilitates optical imaging for identifying few-layer samples, provided by an interferometrically enhanced contrast [2, 29]. However, interferometric effects from this layer could also enhance the measured backreflected harmonic generation [30], leading to an overestimate of the intrinsic nonlinearity of $MoS_2$ (as discussed and measured in Supplementary Material). Therefore, to avoid such effects, we transfer the $MoS_2$ monolayers to a transparent borosilicate glass substrate. The direct dry transfer method described in Ref. [31] is first used to transfer $MoS_2$ to poly(butylene-adipate-co-terephtalate) – (PBAT), which is subsequently placed on the target substrate. The temperature is then raised until melting of the polymer and by using a solvent (chloroform), the polymer is completely removed.

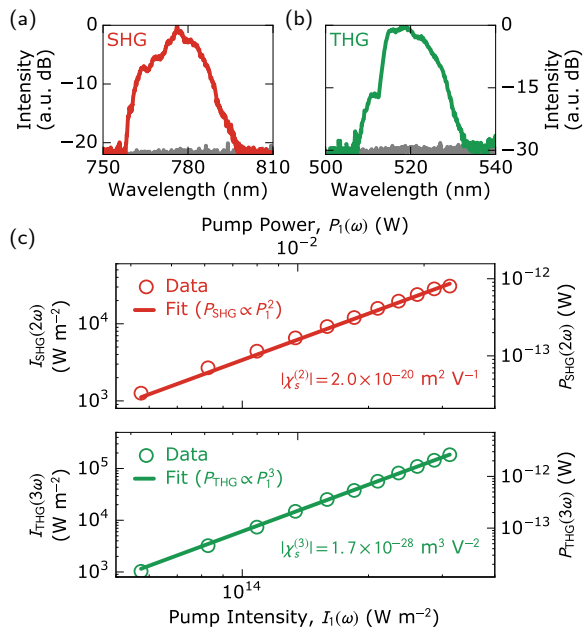

**Figure 3.** Harmonic generation in monolayer $MoS_2$ on glass substrate: optical spectra of (a) second-harmonic and (b) third-harmonic signals (grey lines show the negligible response from the substrate); (c) dependence of generated harmonic intensities upon pump intensity.

The variation in generated harmonic power with pump power for monolayer $MoS_2$ on the glass substrate shows that SHG and THG exhibit the expected quadratic and cubic dependences, respectively [Figure 3(c)]. From (3) and (4), we calculate $|\chi_s^{(2)}| = (2.0 \pm 0.4) \times 10^{-20}$ m² V⁻¹ and $|\chi_s^{(3)}| = (1.7 \pm 0.6) \times 10^{-28}$ m³ V⁻² for monolayer $MoS_2$. The error values are obtained from measurement uncertainties of the terms in Eqns. 3 and 4. Characterization experiments are repeated across 10 different monolayer flakes: we observe 3.7% standard deviation relative to the mean value for the distribution of values of $|\chi_s^{(2)}|$ and 2.9% for $|\chi_s^{(3)}|$, suggesting good repeatability.

Finally, we note that monolayer $MoS_2$ belongs to the $D_{3h}$ point group [18], which enables the polarization-dependence of harmonic generation to be determined from classical nonlinear optical theory [32]. This has already been verified for SHG [18]. We confirmed the expected polarization dependence of THG in a $D_{3h}$ point group crystal for $MoS_2$ using a polarizer: for linearly polarized excitation, the emitted third-harmonic signal is collinearly polarized with the pump wave and as the pump polarization is varied from linear to circular using a quarter waveplate, the intensity of THG is reduced to zero (experimental

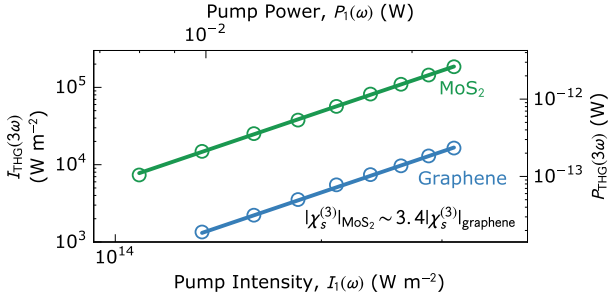

**Figure 4.** Comparison between backward THG versus pump intensity for CVD monolayer MoS$_2$ and graphene.

results and theory are presented in Supplementary Material).

*3.2. Comparison with Graphene*

As the array of available 2D materials grows, it is important to establish their relative nonlinear optical performance. Therefore, we compare the presented results with those for monolayer CVD graphene on a glass substrate, following an identical procedure used for MoS$_2$. This enables a direct comparison of harmonic generation between graphene and MoS$_2$ on the same substrate and in the same setup with 1560 nm excitation (Figure 4). As expected from the inversion symmetry of graphene's atomic structure, SHG is not observed. We do observe THG in graphene, however, from which $|\chi_s^{(3)}| = (0.5 \pm 0.2) \times 10^{-28}$ m$^3$ V$^{-2}$ is computed, suggesting that the third-order nonlinearity of MoS$_2$ is ~3.4 times greater.

This supports earlier observations of stronger saturable absorption, an additional nonlinear effect, in MoS$_2$ compared to graphene [5]. A further benefit of MoS$_2$ is the lack of inversion symmetry, enabling the exploitation of second-order effects (e.g. SHG [16–20] and sum-frequency generation [33]), which are absent in graphene. Monolayer MoS$_2$ could therefore be a superior material than graphene for nonlinear photonic applications at telecommunication wavelengths.

**4. Discussion**

A defining feature of monolayer transition metal dichalcogenides is exciton effects, which can resonantly enhance light-matter interactions. In monolayer MoS$_2$, these excitonic transitions have previously been measured at 1.90 eV (653 nm), 2.05 eV (605 nm) and 2.8 eV (442 nm) [3, 17], labelled A, B and C according to standard nomenclature [34]. Previous SHG studies have reported an enhancement of nonlinear susceptibility values near these resonances: Malard *et al.* measured an off-resonance second-order sheet susceptibility for mechanically exfoliated MoS$_2$ of ~ $1 \times 10^{-20}$ m$^2$ V$^{-1}$, increasing by a factor of ~8 as the SHG wavelength was shifted to overlap with the C exciton [17]. We note good agreement with our measured value of $|\chi_s^{(2)}| = 2 \times 10^{-20}$ m$^2$ V$^{-1}$, for which no resonant enhancement is expected since both pump and second-harmonic are far from excitonic lines. Our 1560 nm pump wavelength is chosen for the potential to realize 2D material-based nonlinear optical devices for telecommunication applications. We note, however, that a stronger nonlinear response could be achieved at other salient pump wavelengths due to excitonic enhancement - e.g. for 1300 nm pumping, the second- and third-harmonic signals are expected to be resonant with the A and C excitons, respectively.

It should be noted that the fabrication method can affect the quality (i.e. defect content) of monolayer MoS$_2$. While mechanically exfoliated samples typically exhibit the highest quality, CVD is a more practical fabrication technique, which is scalable for high-yield production [35]. It is promising that our CVD MoS$_2$ monolayers exhibited similar nonlinear optical susceptibilities to the mechanically exfoliated MoS$_2$ of Malard *et al.* [17]. We also verified this by producing monolayer MoS$_2$ using mechanical exfoliation [3] and comparing THG with that of a CVD sample under identical conditions: less than ~26% variation in the measured susceptibility value was noted. We conclude, therefore, that CVD MoS$_2$ can offer equivalent performance to mechanically exfoliated MoS$_2$ for nonlinear optical applications.

To compare to other literature reports, we relate our measured sheet susceptibilities to an effective bulk nonlinearity: $|\chi_b^{(n)}| = |\chi_s^{(n)}|/h$ where $h$ is the monolayer thickness (0.7 nm for MoS$_2$, 0.335 nm for graphene [29]), yielding $|\chi_{b,\text{MoS}_2}^{(2)}| = 2.9 \times 10^{-11}$ m V$^{-1}$. This is within an order of magnitude of the ~0.6 $\times$ $10^{-11}$ m V$^{-1}$ value at 1560 nm excitation reported by Clark *et al.*, who also tuned their pump wavelength to show a 7× and 5× enhancement in measured nonlinearity for MoS$_2$ on a silica substrate related to the A and B excitons, respectively [20]. Similar order-of-magnitude agreement is also noted with the off-resonance susceptibility value of ~$1 \times 10^{-10}$ m V$^{-1}$, derived by Trolle *et al.* using tight-binding band structure theory including excitonic effects [36].

Our THG measurements are the first characterization of the third-order response of monolayer MoS$_2$ to the best of our knowledge. We note, however, that Wang *et al.* have considered THG from multilayer (>7 layer) MoS$_2$ stacks, deducing an effective third-order susceptibility of ~ $10^{-19}$ m$^2$ V$^{-2}$ [21], which aligns with the bulk value of $|\chi_{b,\text{MoS}_2}^{(3)}| = 2.4 \times 10^{-19}$ m$^2$ V$^{-2}$ that we derive from our sheet nonlinearity measurement. They suggest that enhancement due to band-to-band transitions occurs for all harmonic signals with photon energy exceeding the A exciton transition energy, with greatest enhancement near the A and B exciton. This is supported by their observation that THG is undetectable once the pump is tuned such that the third-harmonic wavelength exceeds ∼660 nm [21].

It is also noteworthy that the generated third-harmonic intensity exceeds that of the second-harmonic. Conventionally, higher-order nonlinear processes are expected to be weaker as more photons are required for the interaction, which occurs with a lower probability. To explain our observation of a stronger THG signal, we note that the 520 nm emission may be enhanced by the edge of the B exciton, and it has also recently been suggested that for sufficiently low pump energies, the SHG signal strength may be decreased due to the energy bands taking part in the nonlinear process being nearly rotationally invariant, with only trigonal warping breaking inversion symmetry [37,38].

Finally, we note that our graphene measurement results in an effective bulk value of $|\chi^{(3)}_{b,\text{graphene}}| = 1.5 \times 10^{-19}$ m$^2$ V$^{-2}$. This was observed for graphene samples we fabricated using both CVD and mechanical exfoliation, and is notably four orders of magnitude weaker than reported by a four-wave mixing study by Hendry *et al.* [11]. It has been noted, however, that a calculation error in Ref. [11] resulted in an overestimate [25]; when corrected, a value of ∼$10^{-19}$ m$^2$ V$^{-2}$ is obtained, in line with fundamental theoretical predictions [25] and also in agreement with our measured value.

## 5. Conclusion

We have comprehensively characterized the magnitude of both the second-order and, for the first time, third-order nonlinear susceptibility of monolayer MoS$_2$ using multiphoton microscopy. The 2D material was treated as a nonlinear polarization sheet, for which sheet susceptibility magnitudes of $|\chi^{(2)}_s| = 2.0 \times 10^{-20}$ m$^2$ V$^{-1}$ and $|\chi^{(3)}_s| = 1.7 \times 10^{-28}$ m$^3$ V$^{-2}$ were calculated from measurements, and direct experimental comparison between graphene and MoS$_2$ showed ∼3.4 times stronger third-order nonlinearity in monolayer MoS$_2$. It was also shown that the nonlinear optical quality of CVD-grown MoS$_2$ was equivalent to mechanically exfoliated MoS$_2$. These results demonstrate opportunities for MoS$_2$ in integrated frequency conversion, nonlinear switching and signal processing, which depend on the magnitude of nonlinear susceptibilities we have characterized within the telecommunications C band.

**Acknowledgments**

We acknowledge funding from the São Paulo Research Foundation (FAPESP), grants 2012/50259-8, 2014/50460-0 and 2015/11779-4, and the Imperial College London Global Engagement Programme. This work is also partially funded by Conselho Nacional de Desenvolvimento Cientfico e Tecnolgico (CNPq) and Fundo Mackenzie de Pesquisa (MackPesquisa). G.E. acknowledges Singapore National Research Foundation for funding under NRF Research Fellowship (NRF-NRFF2011-02) and Medium-Sized Centre Programme. C.P., E.J.R.K. and R.I.W. are supported by fellowships from FAPESP (grant 2015/12734-4), Royal Academy of Engineering and EPSRC, respectively.

# Supplementary Information for

# Characterization of the second- and third-order nonlinear optical susceptibilities of monolayer MoS$_2$ using multiphoton microscopy

R. I. Woodward,[1] R. T. Murray,[1] C. F. Phelan,[2] R. E. P. de Oliveira,[2] T. H. Runcorn,[1] E. J. R. Kelleher,[1] S. Li,[3] E. C. de Oliveira,[2] G. J. M. Fechine,[2] G. Eda,[3] and C. J. S. de Matos[2,*]

[1] Femtosecond Optics Group, Department of Physics, Imperial College London, London, UK

[2] MackGraphe–Graphene and Nanomaterials Research Center, Mackenzie Presbyterian University, São Paulo, Brazil

[3] Centre for Advanced 2D Materials, National University of Singapore, Singapore

[*] cjsdematos@mackenzie.br

# I. DERIVATION OF RELATIONSHIP BETWEEN GENERATED HARMONICS AND NONLINEAR SHEET SUSCEPTIBILITY

When light in air (medium 1) is incident on a material (medium 2), a fraction of the field will be reflected at the interface and the remaining light will be transmitted into it (according to the Fresnel equations[1]). In our case, a MoS$_2$ monolayer is placed on the surface and behaves as a polarization sheet—a layer of radiating dipoles that under intense illumination emits fields at frequencies determined by nonlinear mixing of the pump frequency ($\omega$), a fraction of which is transmitted back into medium 1. Our derivation follows Ref.[2], but is formulated in SI units and extended to consider third-harmonic generation (THG) in addition to second-harmonic generation (SHG).

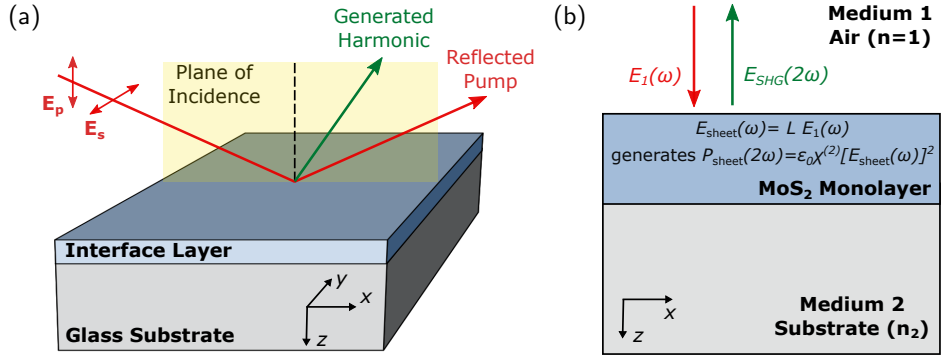

FIG. 1. Illustration showing the treatment of nonlinear sheets at interfaces: (a) generalized nonlinear harmonic generation from a surface, assuming a thin interface layer which acts as a radiating nonlinear polarization sheet under intense illumination; (b) our experimental setup for SHG, showing normal incidence pump light generating a nonlinear polarization wave at frequency $2\omega$ within the MoS$_2$ monolayer at the surface of the substrate. The surface polarization radiates second harmonic light back into the air. Similar treatment applies for THG.

A thorough mathematical treatment of the problem considers the angle at which light meets the surface and the state of optical polarization using the $p$ and $s$ coordinate frame. The general expression for these field components from a sheet polarization at frequency $\omega_s$, emitted back into medium 1 is[2]:

$$\mathbf{E_p}(\omega_s) = i\frac{k_1}{2\epsilon_1 k_{2z}} \left[k_{2z} L_{xx} \mathcal{P}_x(\omega_s)\hat{x} + k_x L_{yy} \mathcal{P}_z(\omega_s)\hat{z}\right] \exp(i\mathbf{k_1} \cdot \mathbf{r} - i\omega_s t) \tag{1a}$$

$$\mathbf{E_s}(\omega_s) = i\frac{1}{2\epsilon_1} \left[k_1 L_{yy} \mathcal{P}_y(\omega_s)\hat{y}\right] \exp(i\mathbf{k_1} \cdot \mathbf{r} - i\omega_s t) \tag{1b}$$

where subscripts $m$ and $h$ indicate the medium ($m = 1, 2$) and axis component ($h = x, y, z$),

respectively, for the wavevector component $k_{mh}$, dielectric constant $\epsilon_m$ and the nonlinear sheet polarization component $\mathcal{P}_h$ induced at the interface. The dielectric constant is related to the material refractive index[3] by $\epsilon_m = \epsilon_0 n_m^2$. $L_{hh}$ is a local field correction factor accounting for the different properties of each medium across the interface, related physically to the well-known transmission Fresnel coefficients[1].

As our experiments are performed at normal incidence to the sample (along $z$), $\mathcal{P}_z$ terms are negligible and $E_p$ and $E_s$ are degenerate ($E_p = E_s = E$), greatly simplifying the mathematics (illustrated in Fig. 1b). The Fresnel transmission factor for normal incidence is therefore $L = L_{xx} = L_{yy} = 2n_1/(n_1 + n_2)$. Additionally, medium 1 is air, resulting in $n_1 = 1$, and medium 2 is glass, which has a relatively small dispersion that we neglect [i.e. $n_2(\omega) = n_2(2\omega) = n_2$]. Therefore, Eqns. 1 are simplified, resulting in an equation for the radiated field from the polarization sheet back into free-space:

$$E_1(\omega_s) = \frac{\omega_s}{2\epsilon_0 c}\left(\frac{2}{1+n_2}\right)\mathcal{P}(\omega_s)\exp(ik_1 z - i\omega_s t) \qquad (2)$$

where

$$\mathcal{P}(\omega_s) = \epsilon_0|\chi_s^{(n)}|E_{\text{sheet}}^n(\omega) = \epsilon_0|\chi_s^{(n)}|\left(\frac{2}{1+n_2}\right)^n E_1^n(\omega) \qquad (3)$$

depends on the specific $n$th order nonlinear effect from a sheet with nonlinear surface susceptibility $\chi_s^{(n)}$ and the coefficient $2/(1+n_2)$ is used to relate the incident field in the interface sheet $E_{\text{sheet}}(\omega)$ to the input field in free space $E_1(\omega)$. Since we reduce this to a scalar problem, the susceptibility tensor is replaced by the complex modulus of the appropriate spatial component.

### A. Second Harmonic Generation

For SHG, we find the backwards SHG amplitude in air by substituting the polarization term $\mathcal{P}(\omega_s = 2\omega) = \epsilon_0|\chi_s^{(2)}|\left(\frac{2}{1+n_2}\right)^2 E_1^2(\omega)$ into Eqn. 2:

$$E_{\text{SHG}}(2\omega) = \frac{(2\omega)}{2c}\left(\frac{2}{1+n_2}\right)^3 |\chi_s^{(2)}|E_1^2(\omega). \qquad (4)$$

Optical intensities are related to field amplitudes[3] by $I = 2\epsilon_0 nc|E|^2$. Using this expression, we rewrite Eqn. 4 in terms of peak intensities:

$$I_{\text{SHG}}(2\omega) = \frac{1}{\epsilon_0}\left[\frac{1}{2c}\left(\frac{2}{1+n_2}\right)^2\right]^3 (2\omega)^2|\chi_s^{(2)}|^2 I_1^2(\omega) = \frac{32|\chi_s^{(2)}|^2\omega^2}{c^3\epsilon_0(1+n_2)^6}I_1^2(\omega) \qquad (5)$$

which need to be converted into time-averaged powers, as measured experimentally. Temporally, the pump light is a train of Gaussian pulses, enabling us to express the peak power $P_{\text{pk}}$ in terms of the measured average power: $P_{\text{pk}} = S P_{\text{av}}/(f t_{\text{fwhm}})$ where the shape factor $S = 0.94$ for Gaussian pulses, $f$ is the pulse repetition frequency and $t_{\text{fwhm}}$ is the FWHM pulse duration. Additionally, our pump light is assumed to be a Gaussian beam in space with $I_{\text{pk}} = 2 P_{\text{pk}}/(\pi r^2)$, leading to the expression:

$$I_{\text{pk}} = \frac{2 P_{\text{av}} S}{\pi r^2 f t_{\text{fwhm}}}. \tag{6}$$

Henceforth, $I$ is used to represent peak intensities (since this determines nonlinear effects) and $P$ to denote the time-averaged powers, which are measured experimentally. As second-order nonlinear polarization is generated in proportion to the square of pump light intensity, the emitted second-harmonic will have Gaussian temporal and spatial profiles but with duration and beam radius reduced by a factor of $\sqrt{2}$.

Finally, we substitute Eqn. 6 into Eqn. 5 for both the average input power $P_1(\omega)$ and the backwards SHG power $P_{\text{SHG}}(2\omega)$, both measured in air, including the duration and beam width correction factor:

$$P_{\text{SHG}}(2\omega) = \frac{16\sqrt{2} S |\chi_s^{(2)}|^2 \omega^2}{c^3 \epsilon_0 f \pi r^2 t_{\text{fwhm}} (1+n_2)^6} P_1^2(\omega). \tag{7}$$

### B. Third Harmonic Generation

We derive an equation relating the THG field to the pump using the same method outlined for SHG, but replacing the polarization term with $\mathcal{P}(\omega_s = 3\omega) = \epsilon_0 |\chi_s^{(3)}| \left(\frac{2}{1+n_2}\right)^3 E_1^3(\omega)$:

$$E_{\text{THG}}(3\omega) = \frac{(3\omega)}{2c} \left(\frac{2}{1+n_2}\right)^4 |\chi^{(3)}| E_1^3(\omega) \tag{8}$$

leading to THG intensity:

$$I_{\text{THG}}(3\omega) = \frac{1}{\epsilon_0^2} \left[\frac{1}{2c}\left(\frac{2}{1+n_2}\right)^2\right]^4 (3\omega)^2 |\chi_s^{(3)}|^2 I_1^3(\omega) = \frac{144 I_1^3(\omega) |\chi_s^{(3)}|^2 \omega^2}{c^4 \epsilon_0^2 (1+n_2)^8}. \tag{9}$$

The cubic dependence of the THG intensity upon pump intensity leads to a greater Gaussian pulse and beam width reduction factor of $\sqrt{3}$, which we include when writing the intensities and powers (using Eqn. 6) to find:

$$P_{\text{THG}}(3\omega) = \frac{64\sqrt{3} S^2 |\chi_s^{(3)}|^2 \omega^2}{c^4 \epsilon_0^2 (f t_{\text{fwhm}} \pi r^2)^2 (1+n_2)^8} P_1^3(\omega). \tag{10}$$

## II. CALIBRATION MEASUREMENTS

As a verification of the accuracy of our experimental system, second- and third-harmonic emission are measured from the surface of bulk samples and used to calculate $|\chi^{(2)}|$ and $|\chi^{(3)}|$. The nonlinear susceptibility of a bulk sample can be related to the nonlinear emission generated at the surface by an incident pump beam using the laws of nonlinear reflection and refraction developed by Bloembergen and Pershan[4]. Eqn. 4.9 in Ref.[4] relates the fields radiated in transmission and reflection by an induced nonlinear polarization at the surface of the bulk crystal to a pump beam at normal incidence. In SI units and neglecting the refractive index dispersion, the reflected field in air is given by[4]:

$$E_{NL} = \frac{\mathcal{P}_{NL}}{\epsilon_0(n_1+n_2)(2n_2)} \quad (11)$$

where $n_2$ is the refractive index of the nonlinear material, $n_1$ is the refractive index of the surrounding medium (i.e. air) and $\mathcal{P}_{NL}$ is the nonlinear polarization. In the case of SHG, $\mathcal{P}_{NL}(2\omega) = \epsilon_0 \chi^{(2)} E(\omega)^2$, where $E(\omega)$ is the pump field transmitted into the bulk sample. Substituting this into Eqn. 11 and expressing it in terms of reflected SHG intensity we find:

$$I(2\omega) = 2n_1 \epsilon_0 c \frac{|\chi^{(2)}|^2 E(\omega)^4}{(n_1+n_2)^2 (2n_2)^2}. \quad (12)$$

Finally, an expression for the second-order susceptibility is found by rearranging Eqn. 12, relating the pump field to the optical intensity $I(\omega)$ in the sample and setting $n_1 = 1$:

$$|\chi^{(2)}| = \left[8n_2^4(1+n_2)^2 \epsilon_0 c \frac{I(2\omega)}{I(\omega)^2}\right]^{\frac{1}{2}}. \quad (13)$$

Starting from the same expression in Ref.[4] and following a similar procedure for THG, the following equation for $\chi^{(3)}$ is obtained:

$$|\chi^{(3)}| = \left[16n_2^5(1+n_2)^2 \epsilon_0^2 c^2 \frac{I(3\omega)}{I(\omega)^3}\right]^{\frac{1}{2}}. \quad (14)$$

The reflected second-harmonic signal from the surface of a ZnS Cleartran prism is measured and a value of $|\chi^{(2)}| = 1.2 \times 10^{-11}$ m V$^{-1}$ is calculated from Eqn. 13. The value for the nonlinear $d_{33}$ coefficient of ZnS given in Shoji et al[5] is $d_{33} = 9 \times 10^{-12}$ m V$^{-1}$, which implies a second order susceptibility $1.8 \times 10^{-11}$ m V$^{-1}$, within 33% of our measured value.

By measuring the third-harmonic signal from the same ZnS material, $|\chi^{(3)}| = 5.1 \times 10^{-21}$ m$^2$ V$^{-2}$ is obtained using Eqn. 14. This is in good agreement (21% difference) with the value of $4.2 \times 10^{-21}$ m$^2$ V$^{-2}$ ($3 \times 10^{-13}$ cm$^3$ erg$^{-1}$) quoted in Weber's *Handbook of Optical Materials*[6].

## III. EXPERIMENTAL MEASUREMENTS OF MOS$_2$ ON SI/SIO$_2$ SUBSTRATE

As discussed in the main text, the SiO$_2$ overlayer on the Si substrate can interferometrically enhance reflected light, which could lead to an overestimate of the intrinsic material nonlinearity of MoS$_2$[7,8]. To quantify this, we perform SHG and THG measurements for monolayer MoS$_2$ on the Si/SiO$_2$ substrate (Fig. 2). The data is well fitted by the equations derived in Section I, from which the sheet susceptibility values are computed as: $|\chi_s^{(2)}| = 2.4 \times 10^{-20}$ m$^2$ V$^{-1}$ and $|\chi_s^{(3)}| = 9.0 \times 10^{-28}$ m$^3$ V$^{-2}$.

Compared to measurements once the MoS$_2$ had been transferred to a transparent glass substrate (see main text), the value of $|\chi_s^{(2)}|$ is slightly increased on the Si/SiO$_2$ substrate, although a significant $\sim 5\times$ enhancement is noted for $|\chi_s^{(3)}|$. This confirms the importance of accounting for possible substrate effects when performing nonlinear characterization measurements.

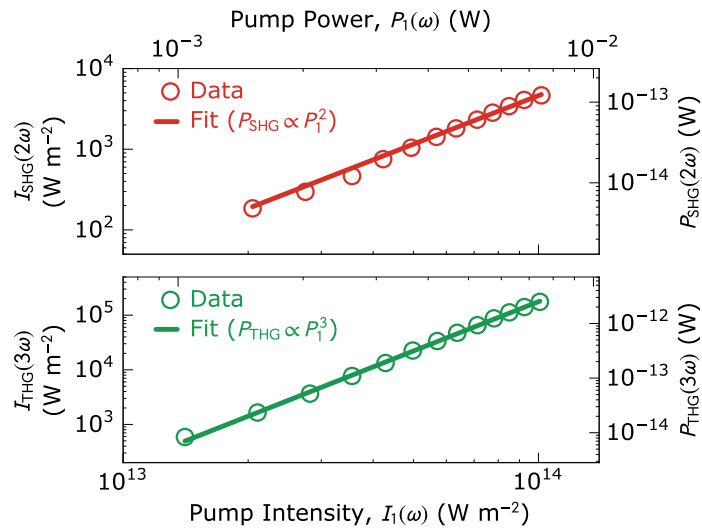

FIG. 2. Dependence of generated harmonic intensities upon pump intensity in monolayer MoS$_2$ on Si/SiO$_2$ substrate.

## IV. POLARIZATION DEPENDENCE OF HARMONIC GENERATION IN MONOLAYER MOS$_2$

As the intensity and state of polarization of generated harmonic light is dependent on the pump polarization and material crystallographic symmetry, polarization-resolved multiphoton microscopy offers a useful route to probing the symmetry property of emerging 2D materials. Recent studies have demonstrated these opportunities using SHG[9]; here, we report the characterisation of THG polarization-dependence in monolayer $MoS_2$.

In THG, the material polarization is related to the pump field by:[10]

$$\mathcal{P}_i(3\omega) = \varepsilon_0 \sum_{jkl} \chi^{(3)}_{ijkl}(3\omega;\omega,\omega,\omega) E_j(\omega) E_k(\omega) E_l(\omega) \tag{15}$$

where $i$, $j$, $k$, and $l$ can each be $\hat{\mathbf{x}}$, $\hat{\mathbf{y}}$ or $\hat{\mathbf{z}}$ directions.

Monolayer $MoS_2$ belongs to the $D_{3h}$ point group[9]. For normally incident light (i.e. neglecting $\hat{\mathbf{z}}$ components), the only non-zero elements of the $\boldsymbol{\chi}^{(3)}$ tensor are:[10] $xxxx = yyyy = xxyy + xyyx + xyxy$ where $xxyy = yyxx$, $xyyx = yxxy$ and $xyxy = yxyx$. The notation $xyyy$ is shorthand for the $\chi^{(3)}_{xyyy}$ element, and we assume $xxyy = xyyx = xyxy$, since these are indistinguishable for THG. To simplify notation, we let $\chi^{(3)}_{xxxx} = \chi$.

For a linearly polarized pump wave, $\boldsymbol{E} = E_x\hat{\mathbf{x}} + E_y\hat{\mathbf{y}}$, where $E_x$ and $E_y$ are real, the material polarization according to Eqn. 15 is: $\boldsymbol{\mathcal{P}}(3\omega) = E_x(E_x^2 + E_y^2)\hat{\mathbf{x}} + E_y(E_x^2 + E_y^2)\hat{\mathbf{y}}$. This shows that THG intensity (related to $|\boldsymbol{\mathcal{P}}(3\omega)|^2$) is independent of the polarization angle for a linearly polarized pump and that the third-harmonic and pump signals will have the same state of polarization. We verified this experimentally using a half-wave plate to rotate the incident polarization and recording the THG polarization state using an analyzer.

In contrast, THG intensity depends strongly on the polarization ellipticity. We explored this experimentally by rotating a quarter-wave plate (QWP) placed in the path of the pump beam: Fig. 3 shows that THG is maximal for linearly polarized light, reducing to zero for circularly polarized light. By rewriting the pump field incident on the sample in terms of the angle of the QWP fast axis relative to the input linear polarization, $\theta$ and substituting this into Eqn. 15, we find $|\boldsymbol{\mathcal{P}}(3\omega)|^2 = \left(\epsilon_0 \chi |E|^3\right)^2 \cos^2(2\theta)$, which is shown to be in good agreement with experimental data in Fig. 3.

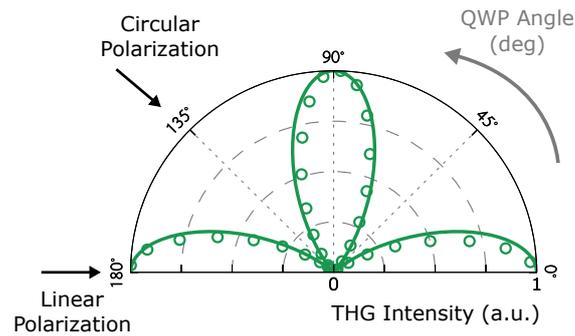

FIG. 3. Polarisation dependence of THG intensity: experimentally measured (circles) dependence as pump polarization is varied using a QWP, in good agreement with the expected behaviour (solid line) for a $D_{3h}$ symmetry class material.